\documentclass[aps,prd]{revtex4}
\usepackage{color}
 \usepackage{amsbsy}
  \usepackage{bm}
 \usepackage{amsfonts}
\usepackage{amssymb}
\usepackage{amsmath}
\usepackage{graphics}
\usepackage{graphicx}
\begin{document}

\title{Regularization of the  Nambu-Jona Lasinio model under a uniform magnetic field and the role of the anomalous
magnetic moments}
\author{R. M. Aguirre}

\affiliation{
\it{Departamento de Fisica, Facultad de Ciencias Exactas}, \\\it{Universidad Nacional de La Plata,} \\
\it{and IFLP, UNLP-CONICET, C.C. 67 (1900) La Plata, Argentina.}}


\begin{abstract}
The vacuum contribution to quark matter under a uniform magnetic
field within the SU(3) version of the Nambu and Jona-Lasinio model
is studied. The standard regularization procedure is examined and
a new prescription is proposed. For this purpose analytic
regularization and a subtraction scheme are used to deal with
divergencies depending on the magnetic intensity. This scheme is
combined with the standard three momentum cutoff recipe, and
reduces to it for vanishing magnetic intensity. Furthermore, the
effects of a direct coupling between the anomalous magnetic
moments of the quarks and  the magnetic field is considered.
Single particle properties as well as bulk thermodynamical
quantities are studied for a configuration of matter found in
neutron stars. A wide range of baryonic densities and magnetic
intensities are examined at zero temperature.
\end{abstract}
\maketitle

\section{Introduction}

The study of dense matter under strong interaction is usually
carried out by employing effective models, due to the intricacies
of the fundamental theory . Within this approach, the Nambu-Jona
Lasinio (NJL) model has shown to be a useful conceptual tool to
tackle different problems. In particular, it has been widely used
to describe quarks interacting with magnetic fields
\cite{MIRANSKY,ANDERSEN,KLEVANSKY,GUSYNIN,EBERT,FERRER,NORONHA,FROLOV,CHATTERJEE,DENKE,
FERRER2,AMAN,PAIS}. Within this versatile description a variety of
issues have been analyzed such as magnetic catalysis, magnetic
oscillations \cite{EBERT}, color superconductivity
\cite{FERRER,NORONHA,CHATTERJEE,AMAN}, chiral density waves
\cite{FROLOV}, vector \cite{DENKE} and tensor \cite{FERRER2}
additional couplings, and quark stars\cite{PAIS}. \\
Multiple efforts have been made to extract physical content from
the vacuum of the strong interaction affected by a magnetic field
\cite{BLAU,GOYAL,ANDERSEN1,COHEN,RUGGERI}. Ref. \cite{BLAU}
provides general expressions for the effective action of a Dirac
field interacting with a magnetic field for intensities greater
than the mass scale. A description based on the chiral sigma
lagrangian has been made in \cite{GOYAL}, \cite{ANDERSEN1} uses
the quark-meson model, whereas in \cite{COHEN} the vacuum
contribution to the magnetization is evaluated in a one-loop
approach to QCD for very intense magnetic fields. Using the chiral
quark model and a Ginzburg-Landau expansion Ref. \cite{RUGGERI}
has found that different treatments of the divergences could yield
important modifications of the phase
diagram.\\
The investigation of this issue has been particularly active
within the NJL model
\cite{KLEVANSKY,GUSYNIN,EBERT,MENEZES,FAYAZ2,ANDERSEN2,AVANCINI0,AVANCINI1,AVANCINI2,FAYAZ,CHAUDHURI},
since the contribution coming from the Dirac sea of quarks is
responsible for the dynamical breaking of the chiral symmetry and
it is a crucial point of the NJL model. There exist several
prescriptions to deal with divergent contributions in the NJL at
zero magnetic field, all of them yield compatible predictions. But
it seems that it is not the case in the presence of an external
magnetic field, as was recently pointed out in
\cite{AVANCINI1,AVANCINI2}. In these references it is mentioned
that the use of smooth form factors instead of a steep cutoff
could change drastically the physical predictions. Particularly
Ref. \cite{AVANCINI2} points out that a reliable regularization
must clearly distinguish between the non-magnetic vacuum
contribution from the magnetic one.  A failure in this point
should be the cause of the inadequate behavior found in different
calculations, such as tachyonic poles in the spectrum of light
mesons and unphysical oscillations in thermodynamical
quantities.\\
Following the analytic regularization in terms of the Hurwitz zeta
function \cite{BLAU}, a residue depending on the squared magnetic
intensity is found in \cite{EBERT}. To dispose of this
singularity, the authors propose a wavefunction renormalization by
associating it to the pure magnetic contribution to the energy
density. To deal with the divergencies in the thermodynamic
potential, in \cite{MENEZES,FAYAZ2} the pressure at zero baryonic
density and finite $B$ is subtracted and added, in the former case
still exhibiting the undesirable divergency and in the latter one
it has been regularized in the 3-momentum cutoff scheme. While in
the calculations of \cite{FAYAZ} a softening regulator is used to
analyze the effects of the AMM on the quark matter phase diagram,
in \cite{CHAUDHURI} a step function in momentum space is used with
the same purpose.

An interesting aspect to be taken into account for Dirac particles
in a magnetic field is the discrepancy of the gyromagnetic factor
from the ideal value 2. This can happen in a quasiparticle scheme
where the effects of some interactions give rise to the anomalous
magnetic moments (AMM)\cite{SINGH,BICUDO,FERRER2}. As a reference
one can take the prediction of the non relativistic constituent
quark model for the magnetic moments of the light quarks. In order
to adjust the experimental values of the proton and neutron
magnetic moments, the gyromagnetic ratios $\tilde{g}_u=2
\mu_u/\mu_N=3.7$,
$\tilde{g}_d=2 \mu_d/\mu_N=-1.94$ are obtained within this approach.\\
The appearance of AMM is closely related to the breakdown of the
chiral symmetry. For this reason the NJL model with zero current
quark mass have been used to study the origin of the AMM
\cite{FERRER2,SINGH,BICUDO}.
 To analyze the feasibility of the dynamical generation of the AMM, in
\cite{SINGH} a one loop correction to the electromagnetic vertex
is evaluated within the one flavor NJL model, obtaining
\[ \tilde{g}\approx \frac{2 M_p}{M}\frac{e_q}{e}\left[1-(M/\Lambda)^2 \log(\Lambda/M)\right]. \]
In a more sophisticated treatment they obtain $\tilde{g}_u\approx
3.72, \; \tilde{g}_d\approx -1.86$, by choosing adequately the
constituent quark masses.\\
In the approach of \cite{BICUDO} the AMM is extracted from the low
momentum electromagnetic current written in terms of the kernel of
the Ward identities. Assuming a four momentum cutoff, they find
zero AMM in a one flavor NJL model. However,  by using the two
flavor version, the authors obtain $\tilde{g}_u\approx 3.813, \;
\tilde{g}_d\approx -1.929$, which differ from the phenomenological
expectations by less than $1 \%$. Furthermore, in the same work an
schematic confining potential for only one flavor is considered.
By taking a constituent quark mass $M=330$ MeV, typical of the NJL
model, the magnitude of the AMM
predicted is as large as 0.15.\\
Another point of view is developed in \cite{FERRER2}, where the
one flavor NJL model is supplemented with a four fermion tensor
interaction, which induces a condensate in the $\gamma^1\,
\gamma^2$ channel. In this case the intrinsic relation between the
 constituent quark mass and its AMM is explicitly exposed, since
the vacuum condensate which breaks the chiral symmetry is also
responsible for the occurrence of nonzero AMM. As a consequence,
the AMM
has a non-perturbative dependence on the magnetic intensity.\\
The necessity of the AMM of the quarks has been emphasized in
\cite{MEKHFI} in the context of the Karl-Sehgal formula, which
relates baryonic properties with the spin configuration of the
quarks composing them. By stating the dynamical independency of
the axial and tensorial quark contributions to the baryonic
intrinsic magnetism, the AMM of the quarks are proposed as the
parameters that distinguish between them. Resorting to sound
arguments, the author propose  $a_u=a_d\approx 0.38$, $a_s\approx
0.2-0.38$ as significative values for the AMM for the lightest
flavors.

Other investigations have focused on the consequences of a linear
coupling between the AMM of the quarks and an external magnetic
field \cite{FAYAZ,CHAUDHURI,GONZALEZ}. For instance, \cite{FAYAZ}
analyze the phase diagram of the NJL at finite temperature, with
special emphasis on a possible chiral restoration due to the
non-zero AMM. Furthermore, the possibility of a non linear
coupling of the AMM of the quarks is considered. In this model the
AMM is related to the quantum correction to the electrodynamic
vertex, and a non-perturbative dependence on the magnetic field is
introduced through the effective constituent quark mass.  The
influence of the AMM on the structure of the lightest scalar
mesons is analyzed in \cite{CHAUDHURI}, while \cite{GONZALEZ} is
devoted to study their effects on neutral and beta stable quark
matter within a bag model.

The aim of the present work is to study the effects of a uniform
magnetic field on the properties of matter of quarks that have
acquired AMM. In particular we focus on the vacuum effects and we
perform an analytical regularization of the NJL that matches the
standard three-momentum cutoff scheme for vanishing magnetic
intensity. For this purpose, a fermion propagator is used which
includes the anomalous magnetic moments and the full interaction
with the external magnetic field \cite{AGUIRRE1,AGUIRRE2}. This
propagator has been used to evaluate meson properties
\cite{AGUIRRE2, AGUIRR3}, and the effect of the AMM within the NJL
model \cite{CHAUDHURI}. Previous investigations have considered
quarks with AMM within this framework \cite{FAYAZ,CHAUDHURI}, but
the divergent one-dimensional integrals were treated with a
momentum cutoff which depends on the magnetic intensity. In the
case of \cite{FAYAZ} the cutoff parameter $\Lambda$ is inspired by
a covariant 4-momentum scheme $E_{n \, s}(p,B)<\Lambda$, where
$E_{n \, s}(p,B)$ represents the energy of the $n$-th Landau level
with spin projection $s$ along the direction of the uniform
magnetic field. Ref. \cite{CHAUDHURI}, instead, uses a 3-momentum
cutoff
$p<\sqrt{\Lambda^2+M^2-E_{n \, s}(0,B)^2}$.\\
In the present work a 3-flavor version of the NJL is used. Most of
the references just cited, in particular those corresponding to
the study of AMM, use the two-flavor formulation. Thus we provide
here an insight on the dynamics of the strange degree of freedom.
This is particularly useful for applications to astrophysical
studies, as for instance the final stage of neutron stars, where
quark matter is electrically neutral and it is in equilibrium
against weak decay.

This work is organized as follows. In the next section a summary
of the NJL model is presented and a set of prescriptions to deal
with the divergent contributions of the Dirac sea of quarks with
AMM is proposed.  Some numerical results are discussed in Sec.
III, and the last section is devoted to drawing the conclusions.

\section{Effects of the AMM on the vacuum properties in the NJL model}

The SU(3) NJL model extended with an AMM term  has the Lagrangian
density

\[
{\cal L}_{NJL}=\bar{\psi}\left(i
\slash\!\!\!\!D-M_0-\frac{1}{2}\kappa \, \sigma_{\mu\,\nu}
F^{\mu\,\nu}\right)\psi+G\left[\left(\bar{\psi}\lambda_a
\psi\right)^2+\left(\bar{\psi} i \gamma^5 \lambda_a
\psi\right)^2\right]-K \left[\text
{det}\,\bar{\psi}(1+\gamma^5)\psi+\text
{det}\,\bar{\psi}(1-\gamma^5)\psi\right],
\]

where a summation over color and flavor is implicit and the
current mass matrix $M_0=\text{diag}(M_{0 u},M_{0 d},M_{0 s})$
breaks explicitly the chiral symmetry and the covariant derivative
$D^\mu=\partial^\mu-i Q e A^\mu/3$ takes account of the uniform
magnetic field, with $Q=\text{ diag}(2,-1,-1)$. The AMM are
displayed in the matrix $\kappa=\text{
diag}(\kappa_u,\kappa_d,\kappa_s)$ and the definition $\sigma_{\mu
\nu}=i[\gamma_\mu,\gamma_nu]/2$ is used. In the following only the
zero temperature case is considered.

Due to the presence of a vacuum condensate the quark field
acquires an enlarged constituent  mass, a process that in the
usual Hartree approach is described by
\begin{equation}
M_i=M_{0 i}- 4 G  \langle\bar{\psi}_i \psi_i\rangle+ 2 K
\langle\bar{\psi}_j \psi_j\rangle\langle\bar{\psi}_k
\psi_k\rangle, \label{EffMass}
\end{equation}
for $i\neq j \neq k$.

By using standard techniques \cite{KLEVANSKY,KUNIHIRO,VOGL} one
finds the grand partition function per unit volume $\Omega$,
\begin{equation}
\Omega=\sum_f \left[N_c \, T_f+2 G
\langle\bar{\psi}_f\psi_f\rangle^2\right]-4
K\langle\bar{\psi}_u\psi_u\rangle\langle\bar{\psi}_d\psi_d\rangle\langle\bar{\psi}_s\psi_s\rangle.
\label{GrandPotential}
\end{equation}
In the limit of zero temperature the first term between square
brackets can be decomposed as $T_f= T_{0 f}-\mu_f \,n_f $, where
$n_f$ stands for the particle number density for a given flavor
$f$, and the Lagrange multipliers $\mu_f$ manifest the
simultaneous conservation of the electric charge and the baryonic
number. Alternatively, the kinetic contribution $T_{0 f}$ can be
expressed as $T_{0 f}=\langle\bar{\psi}_f\,i
\gamma^0\partial_0\,\psi_f\rangle$  \cite{KLEVANSKY} and
eventually can be evaluated in terms of the single particle Green
function.
\\
Both quantities $\langle\bar{\psi}_k\psi_k\rangle$ and $\langle
\bar{\psi}_k i\gamma^0\partial_0 \psi_k\rangle$ are ultraviolet
divergent and need to be interpreted adequately. There are several
standard recipes within the NJL model, such as the non-covariant
3-momentum cutoff and the Lorentz invariant procedures of
Pauli-Villars and 4-momentum cutoff. In the present work a
regularization procedure is used which reduces to the 3-momentum
cutoff at  vanishing magnetic field. For this purpose a fermionic
propagator is used corresponding to an effective quark with
constituent mass and interacting with an uniform magnetic field
through the electric charge and the AMM. It  has been deduced for
positively charged fermions in \cite{AGUIRRE1} within the real
time formalism of the thermal field theory Thermo Field Dynamics
\cite{LANDSMAN}. For the sake of completeness the results
corresponding to zero temperature are transcribed here
\begin{equation}
G_f(x',x)=e^{i \Phi } \int \frac{d^4 p}{(2 \pi)^4} e^{-i
p^\mu\,(x'_\mu-x_\mu)} \left[ G_{f 0}(p)+e^{-p_\bot^2/\beta_f}
\sum_{n,s}(-1)^n  G_{f n s} (p)\right] \label{PropP}
\end{equation}
where
\begin{eqnarray}
G_{f 0}(p)&=&e^{-p_\bot^2/\beta_f}\left( \not \! u+M_f-K_f\right)
\left(1+i \gamma^1 \gamma^2\right) \; \Xi_{0 s},\label{LLL}
\end{eqnarray}
\begin{eqnarray}
G_{f n s}(p)&=&\frac{\Delta_n+s M_f}{2 \Delta_n}\Big\{( \not \!
u-K_f+s \Delta_n) \left(1+i \gamma^1 \gamma^2\right) L_n(2
p_\bot^2/\beta_f)-( \not \! u+K_f-s \Delta_n)
\nonumber\\
&&\times \left(1-i \gamma^1 \gamma^2\right) \frac{s
\Delta_n-M_f}{s \Delta_n+M_f} L_{n-1}(2 p_\bot^2/\beta_f)+ \left(
\not \! u\, i \gamma^1 \gamma^2+ s \Delta_n-K_f\right) \not \! v
\frac{s \Delta_n- M_f}{p_\bot^2}
\nonumber \\
&&\times \left[ L_n(2 p_\bot^2/\beta_f)-L_{n-1}(2
p_\bot^2/\beta_f)\right]\Big\}
\; \Xi_{n s},  \\
\Xi_{n s}&=&\frac{1}{p_0^2-E_{f n s}^2+i\epsilon}+2
\pi\,i\,n_F(p_0)\,\delta(p_0^2-E_{f n s}^2). \label{DecomP}
\end{eqnarray}

A similar expansion holds for negatively charged particles. In
these expressions the index $s=\pm 1$ describes the spin
projection on the direction of the uniform magnetic field. Eq.
(\ref{LLL}) propagates the lowest Landau level with the unique
projection $s=1$ for the $u$ flavor and $s=-1$ for the $s, d$
cases. The sum over the index $n\geq 1$ takes account of the
higher Landau levels, and the following notation is used
$\beta_f=e |Q_f| B/3$, $K_f=\kappa_f\,B$, $\not \! u=p_0
\gamma^0-p_z \gamma^3$, $\not \! v=-p_x \, \gamma^1-p_y\,
\gamma^2$, $p_\bot^2=p_x^2+p_y^2$, $L_m$ stands for the Laguerre
polynomial of order $m$, and
\begin{eqnarray}
E_{f n s}&=&\sqrt{p_z^2+(\Delta_n-s\,K_f)^2}\nonumber \\
\Delta_n&=&\sqrt{M_f^2+2 n \beta_f} \nonumber
\end{eqnarray}
Furthermore, $n_F(p_0)=\Theta\left(\mu_f-E_{f n s}\right)$ stands
for the canonical statistical distribution function for fermions
in thermodynamical equilibrium. Finally, the phase factor
$\Phi=\beta_f(x+x')(y'-y)/2$ embodies the gauge fixing.

Using the propagator of Eq. (\ref{PropP}) the quark condensates
and the kinetic contributions are evaluated as \cite{KLEVANSKY}
\begin{eqnarray}
\langle\bar{\psi}_f\,\psi_f\rangle&= &-i\;\lim_{t'\rightarrow
t^+}\,\text{Tr}\{G_f(t,\vec{r},t',\vec{r})\}, \label{Cond}
 \\
\langle\bar{\psi}_f\,i
\gamma^0\partial_0\,\psi_f\rangle&=&-i\;\lim_{t'\rightarrow
t^+}\,\text{Tr}\left\{i \gamma^0 \frac{\partial}{\partial
t}\,G_f(t,\vec{r},t',\vec{r})\right\}, \label{Kinetic}
\end{eqnarray}
together with
\begin{eqnarray}
\langle\bar{\psi}_f\,
\gamma^\nu\,\psi_f\rangle&=-i\;\lim_{t'\rightarrow
t^+}\,\text{Tr}\{\gamma^\nu\,G_f(t,\vec{r},t',\vec{r})\}.
\label{BNumb}
\end{eqnarray}
The principle of thermodynamical consistency can be imposed
through the relation $0=\partial \Omega/\partial M_{0 f}$
\cite{BUBALLA}.\\
 As already mentioned, these  quantities have
divergent vacuum contributions. In the Appendix a regularization
procedure is applied that ensures null vacuum contributions at
zero magnetic intensity. This requisite is used to match the
3-momentum cutoff procedure, by simply adding the standard
expressions in terms of the cutoff parameter $\Lambda$. Thus at
the regularization point these quantities reduce to the commonly
used vacuum values. But for any other conditions, finite
contributions depending on the density and
the magnetic intensity are extracted from the vacuum.\\
Eq. (\ref{BNumb}) represents the density of baryonic current,
which for infinite homogeneous matter has zero vacuum value.\\
In the Appendix the derivation of the regularized Dirac sea terms
of Eq. (\ref{Kinetic}) is shown. That expression reduces to
\begin{eqnarray}
\langle\bar{\psi}_f\, i \gamma^0
\partial_0 \psi_f\rangle^{\text{D}}-\langle\bar{\psi}_f\, i \gamma^0
\partial_0 \psi_f\rangle_{NJL}&\rightarrow&-\frac{N_c \beta_f^2}{2
\pi^2}\left[
\zeta'\left(-1,x\right)-\frac{1}{2}\left(x^2-x+\frac{1}{6}\right)
\ln(x)+\frac{\tilde{x}^2}{4}\right]+\frac{N_c \beta_f^2}{48
\pi^2}\nonumber \\&+&\frac{N_c \beta_f^2}{4
\pi^2}\left(x^2+x+\frac{1}{6}\right)
\ln\left(\frac{\nu}{M_f^2}\right) - \frac{N_c \beta_f^2}{4
\pi^2}\left(3  \tilde{x}^2-\tilde{x}+\frac{1}{6}\right)
\ln\left(\frac{\nu}{\tilde{M}_f^2}\right) \label{Comparison}
\end{eqnarray}
for $\kappa_f \rightarrow 0$. The notation $x=M_f^2/(2 \beta_f)$
and $\tilde{x}=\tilde{M}_f^2/(2 \beta_f)$ has been used, where
$\tilde{M}_f$ stands for the constituent mass at finite baryonic
density and $B=0$.  In this form, it can be compared with previous
results, as for instance \cite{EBERT}
\[-\frac{N_c \beta^2}{2
\pi^2}\left[ \zeta'\left(-1,x\right)-\frac{1}{2}\left(x^2-x\right)
\ln(x)+\frac{x^2}{4}\right] \]

The first term of Eq. (\ref{Comparison}) resembles the last
equation. However, they differ in two points. First, the
polynomial multiplying the logarithm has an extra term, which
comes from the definition of $\zeta(-1,x)$. Furthermore in the
last term between square brackets the quantities $\tilde{M}$ and
$M$ are taken as identical. The remaining terms of Eq.
(\ref{Comparison}) are missing in the mentioned approach. The
difference can be minimized by choosing $\nu=M_f^2$. In such case,
the third term of Eq.(\ref{Comparison}) becomes null, and the last
one would also be zero if one identifies $\tilde{M}=M$. For this
reason we adopt in the following $\nu=M_f^2$, but the distinction
between $M_f$ and $\tilde{M}_f$ will be kept.

Furthermore, Eqs. (\ref{Cond})-(\ref{BNumb}) receives finite
contributions from the Fermi sea
\begin{eqnarray}
\langle\bar{\psi}_f\psi_f\rangle^{\text{F}}&=&\frac{N_c}{2
\pi^2}\beta_f M_f {\sum_{n,s}}^\prime\frac{\Delta_n-s
K_f}{\Delta_n} \ln\left(\frac{\mu_f+p_{f n s}}{\Delta_n-s
K_f}\right),\label{FermiCondensate}
\end{eqnarray}
\begin{eqnarray}
\langle\bar{\psi}_f\, i \gamma^0
\partial_0 \psi_f\rangle^{\text{F}}&=&\frac{N_c}{2
\pi^2}{\sum_{n,s}}^\prime\left[\mu_f\, p_{f n s}+\left(\Delta_n-s
 K_f\right)^2\ln\left(\frac{\mu_f+p_{f n
s}}{\Delta_n-s K_f}\right)\right], \label{FermiEnergy}
\end{eqnarray}
\begin{eqnarray}
n^\nu_f=\langle\bar{\psi}_f \gamma^\nu
\psi_f\rangle^{\text{F}}&=&\delta_{\nu 0}\frac{N_c}{6
\pi^2}\beta_f {\sum_{n,s}}^\prime p_{f n s}, \label{BNumbF}
\end{eqnarray}
where the primed sum indicates that for $n=0$ only one spin
projection must be considered as explained previously. The highest
occupied Landau level N is defined by the condition
$\mu_f^2-(\Delta_N-s K_f)^2\geq 0$. The Lagrange multipliers
$\mu_f$ are determined by the conserved charges, and $p_{f n
s}=\sqrt{\mu_f^2-(\Delta_N-s K_f)^2}$.

 The magnetization per unit volume is given by the equation
${\cal M}=-\partial \Omega/\partial B$, which can be simplified by
using the stationary point conditions \cite{BRODERICK,AVANCINI0},
\[
{\cal M}=\frac{N_c}{8\pi^2 B}\sum_f\left({\cal D}_f+2\,\beta_f
{\sum_{n, s}}^\prime {\cal F}_{n s}\right),
\]
where
\begin{eqnarray}
{\cal D}_f&=&8 \beta_f^2\, \zeta'\left(-1,\omega_f\right)-2
\beta_f \left[M_f^2-
K_f^2\right]\,\ln\left(\frac{\Gamma\left(\omega_f\right)}{\sqrt{2\pi}}\right)-
\left[2 K_f^2\left( K_f^2 + M_f^2 - \beta_f
\right)-\beta_f\left(M_f^2
+K_f^2-\frac{\beta_f}{3}\right)\right]\,\ln\left(\frac{M_f}{2\beta_f}\right)
\nonumber\\
&+&2 \beta_f\left(M_f+ K_f\right)\left(M_f+3
K_f\right)\ln\left(1+\frac{K_f}{M_f}\right)-4 \left[ K_f^2\left(
K_f^2 + \tilde{M}_f^2 \right)+\beta_f\left(2 \tilde{M}_f
K_f+\frac{\beta_f}{3}\right)\right]\,\ln\left(\frac{\tilde{M}_f}{2\beta_f}\right)\nonumber\\
&-&\frac{1}{2}\left(M_f^2-K_f^2\right)^2-\frac{\beta_f^2}{3}-2
\beta_f  K_f \left( 2 \tilde{M}_f-M_f\right),
\end{eqnarray}
\[
{\cal F}_{f n s}=\mu_f\, p_{f n s}-\left(s \Delta_n-
K_f\right)\left(2 s \Delta_n-3  K_f- s \frac{M_f}{\Delta_n}\right)
\ln\left(\frac{\mu_f+p_{f n s}}{|s \Delta_n- K_f|}\right).
\]
The pressure and energy density are given by the canonical
results, $P=-\Omega,\, E/V=\sum_f \mu_f n_f-P$ and the transversal
component of the stress tensor is defined as
$P_\bot=P-{\cal M} B$.\\
 Following a common practice, the quantum corrections to the leptonic properties are neglected,
as well as the effects of their AMM, so that they contribute with
\begin{eqnarray}
n_l&=&\frac{\beta_l}{2\pi^2}{\sum_{n, s}}^\prime p_{l n s} ,\\
P_l&=&\frac{\beta_l}{4\pi^2}{\sum_{n, s}}^\prime\left[\mu_l p_{l n
s}-\left(m_l^2+2 n \beta_l\right)\ln\left(\frac{\mu_l+p_{l n
s}}{\sqrt{m_l^2+2 n \beta_l}}\right)\right],
\\
{\cal M}_l&=&\frac{\beta_l}{4 \pi^2}{\sum_{n, s}}^\prime
\left[\left(m_l^2+4 \beta_l n\right) \ln\left(\frac{\mu_l+p_{l n
s}}{\sqrt{m_l^2+2 n \beta_l}}\right)-\mu_l p_{l n s}\right],
\end{eqnarray}
to the particle number density, pressure and magnetization,
respectively. The definition $p_{l n s}=\sqrt{\mu_l^2-m_l^2-2 n
\beta_l}$ is used.

\section{Results and discussion}

In this section the effects of the AMM of the quarks are studied
for the case of electrically neutral matter and in equilibrium
against weak decay, so it is necessary to include leptons in this
description. The leptons get a chemical potential $\mu_l$
associated with the local conservation of the electric charge,

As it is usual, the conditions for the conservation of the
baryonic charge $n_B=(n_u+n_d+n_s)/3$ and electric neutrality $2
n_u-n_d-n_s-3 n_l=0$  are imposed. The baryonic density of quarks
is given by Eq. (\ref{BNumbF}).

In the present  calculations the following NJL parameters are
used: $M_{u 0}=M_{d 0}=5.5$ MeV, $M_{s 0}=135.7$ MeV,
$\Lambda=631.4$ MeV, $G=1.835/\Lambda^2, \, K=9.29/\Lambda^5$
\cite{KUNIHIRO}. For the total magnetic moments the prescription
$\mu_u=\left(4
\mu_p+\mu_n\right)/5,\;\mu_d=\left(\mu_p+\mu_n\right)/5,\;\mu_s=\mu_\Lambda$
of the constituent quark model is adopted. Taking the experimental
values of the baryonic magnetic moments together with constituent
masses estimated within the same framework $M_u=M_d=363$ MeV and
$M_s=538$ MeV the following AMM are obtained
$\kappa_u=0.074,\;\kappa_d=0.127,\;\kappa_s=0.053$ in units of the
nuclear magneton, this set will be denoted in the following as
AMM1. The values so obtained are small in comparison with other
predictions \cite{BICUDO,MEKHFI}, therefore the alternative set
$\kappa_u=\kappa_d=0.38,\;\kappa_s=0.25$ is also considered. It is
compatible with the results of \cite{MEKHFI}, and
will be recognized as set AMM2.\\
 The range of magnetic intensities
studied $10^{15} \leq B\leq  10^{19}$ G greatly exceeds the
phenomenology of strongly magnetized compact stars.

As a first step, different prescriptions for the regularization of
the NJL immersed in a uniform magnetic field are considered. A
comparison between the present approach and the commonly used
procedure  as described for instance in \cite{EBERT}, is made
here. In the following the last approach is referred as case C,
while the label AMM0 is used for the results of this work when the
AMM are zero. In Fig. 1 the constituent quark masses at zero
baryonic density are shown as a function of $B$, the wide range of
magnetic intensities has the purpose of comparison with previously
published works. To appreciate the low intensity behavior a small
figure is inserted in the upper panel, restricted to $B < 10^{19}$
G. All the approaches agree to predict increasing quark masses,
but the rate of growth is always greater in the case C. A
comparison between this case and the AMM0 one shows that the
difference is accentuated as the magnetic intensity grows, and it
is considerable at extreme intensities. However, a regime of
qualitative coincidence is found for $B < 2 \times
10^{18}$ G. \\
It can be appreciated that the increase of the AMM has opposite
effects on the $u$ flavor as compared to the $d, s$ cases. A
progressive increase in the magnitude of the AMM enhances the rate
of growth for $M_u$, while it attenuates the changes in $M_d,
M_s$. Due to the smallness of the set AMM1 their results are
closer to the AMM0 than to the AMM2. \\
Calculations of the light quark masses at finite temperature
including AMM have been presented in \cite{CHAUDHURI}. A
comparison with these results is risky because they have been
obtained in different conditions, i.e. equal particle number of
$u$ and $d$ flavors and finite temperature. However, the curves
for temperatures below $T=100$ MeV seems to behave similarly. In
Fig. 4 of this reference the variation of the quark mass for $B <
10^{20} G$ shows quick oscillations around a decreasing mean value
when AMM are included, and a slightly decreasing trend is obtained
for zero AMM. In contrast, \cite{FAYAZ} found and almost
monotonous increasing trend at zero temperature, and to the
greater AMM (set $\kappa_1$) corresponds a weaker growth. \\
The influence of the regularization scheme on the vacuum
contribution to the energy density is examined in Fig. 2. In  both
AMM0 and C approaches a decreasing energy is expected within the
range considered here. However in the first case the variation is
only of a few MeV, while it exceeds $10$ MeV for intensities
slightly above $10^{19}$ G in the last instance.\\
 In conclusion, one can say that there is a qualitative
agreement between these procedures in the low magnetic intensity
regime, but the discrepancies become important for $B> 10^{19}$ G.

In Fig.3 the density dependence of the constituent masses is shown
at fixed intensity $B=10^{19}$ G. The figure extends up to
baryonic densities $n=7 n_0$,  a density which is feasible in the
core of magnetars. The reference value $n_0=0.15$ fm$^{-3}$
corresponds to the saturation density of nuclear matter.  A
monotonously decreasing behavior is obtained for all the flavors.
In the case of $M_u$ there is an almost linear trend at low
density till the point $R \approx 1.3$ where the first excited
Landau level starts to be occupied. Here a noticeable change of
slope takes place. The curve for $M_s$ shows a shoulder shape,
after a plateau for $3 < n/n_0 < 4.5$ a change of slope together
with an inflexion point occurs around $n/n_0=4.2$. At this density
the strange quark comes out to the Fermi sea. The effect of the
AMM is almost indistinguishable at the scale shown, but the
numerical increments are at most of 10 MeV for the u flavor and
around 0.1 MeV for the $s$ quark. The influence of the AMM on the
masses of the quarks decreases quickly with the magnetic
intensity, so that for $B <5 \times 10^{18}$ G all the corrections
diminish about 30$\%$.

To take a view of  the density effects in the interior of a
magnetar a model of the variation of the magnetic field with the
density is considered \cite{PAL}. It is given in terms of the
ratio $R=n/n_0$ by the formula
\begin{equation}
B(n)=B_s+B_0 \left[1-\exp(-\beta R^\gamma)\right], \label{BModel}
\end{equation}
where $B_s=10^{15}$ G is the intensity on the star surface, and
the remaining parameters have been chosen as $B_0=5 \times
10^{18}$ G, $\beta=0.01, \; \gamma=3$. The maximum strength
$5\times 10^{18}$ G corresponds to asymptotic high densities and
could not be reached in a realistic description, hence by the
facts just discussed one could expect that the effective quark
masses do not manifest the details of the model. For this reason
some thermodynamical quantities are examined. The thermodynamical
pressure at zero temperature as a function of the baryonic density
is exhibited in Fig. 4(a) for a range which covers from the
surface to medium depths of a typical neutron star. In the same
figure the calculations corresponding to fixed $B=10^{17}$ G and
the three sets of the AMM are included. Only small differences are
found and at the scale shown all the results seem to coincide. A
regime of thermodynamical instability extends for low densities
until $R \approx 1.5$ giving rise to the hadronization process.
For higher densities $R > 3$ the pressure grows almost linearly.
The lower panel, Fig. 4(b), is devoted to the energy per particle
as a function of the density. In this figure a contrast between
the model of density dependent intensity of Eq. (\ref{BModel})
(case D)and the results of the set AMM1 at different intensities
$B$ is presented. For a given density an increase of the magnetic
intensity lessens the energy per particle within the set AMM1. As
expected, the outcome of the case D lies between the curves of
$10^{17}$ G and $5\times 10^{18}$ G of the parametrization AMM1.

The abundance of particles relative to the total number of quarks
is displayed in Fig. 5 as a function of the baryonic density for
the fixed intensity $B=5 \times 10^{18}$ G. There are no
appreciable discrepancies between the different treatments. The
inclusion of the AMM produce a slight shift to lower density of
the rise of the strange quark population. The population of the
strange flavor shows sudden changes of slope, which are noticeable
for the sets AMM0 and AMM1, coincident with the occupation of a
higher Landau level.

The magnetization is a measure of the response of the system to
the magnetic excitation, it is shown in Fig. 6 as a function of
the density.  Since this is a very small quantity, the results are
scaled with the proton charge, which is appropriate for the range
of intensities examined here. In the present case the
magnetization receives contributions from the electrons and from
the three quark flavors in a proportion determined by the local
charge neutrality condition. Different curves corresponding to the
constant intensities $B=10^{17}$ G and $B=5 \times 10^{18}$ G and
the three sets of AMM are displayed. The bottom of this figure is
occupied by the lowest intensity results. Because of their similar
behavior, the high frequency and the small amplitude of their
oscillations, the results of the three parametrizations coalesce
into a band. In such conditions the system is essentially
diamagnetic for almost all densities. Four different regimes can
be clearly distinguished, the limit of zero density with vanishing
values of $\cal{M}$ but a steep negative slope. The second one
corresponds to low density and extends approximately over the
thermodynamic instability region. Here the mean value of the
magnetization takes medium values $|{\cal M}/e|< 0.08$ fm$^{-2}$.
For intermediate densities up to the threshold of arising into the
Fermi sea of the strange flavor, where the mean of $|{\cal M}/e|$
has their lowest values and remains almost stationary. Finally the
high density domain starts with a sudden decrease of the
magnetization, which stabilizes asymptotically around the value
${\cal M}/e \approx -0.14$ fm$^{-2}$. When the magnetic intensity
is increased to $B=5 \times 10^{18}$ the pattern just described is
kept but other significative differences become apparent. The
oscillations have greater amplitudes and do not show a
quasi-periodic distribution. This is a manifestation of a coherent
dynamics, more favorable at higher intensity  due to the smaller
number of accessible Landau levels. Furthermore, in the medium
density regime the system is definitely paramagnetic while for the
higher densities is only moderately diamagnetic. For the model of
variable intensity of Eq. (\ref{BModel}) the magnetization follows
approximately the behavior of the case $B=10^{17}$ G until $R
\approx 2$ where it increases abruptly developing quasi periodic
oscillations of decreasing frequency and increasing amplitude. In
the high density domain it acquires features similar to the case
$B=5 \times 10^{18}$ G.

In Fig. 1 the effects of the magnetic intensity on the quark
masses in vacuum  has been shown.  In order to study how the
density influence the magnetic dependence, a detail of the results
obtained for these masses for non-zero baryonic number is
presented in Fig 7. The values chosen for the density, $R=4$ and
$R=7$, correspond to situations where quark matter is stable and
the s quark is only virtual or is able to exist in the Fermi
shell, respectively. A well distinguishable behavior is obtained
for the three flavors. The light quarks show a monotonous behavior
for the selected densities and the full range of intensities. An
increase in the magnitude of the AMM implies an increase for $M_u$
independently of the density chosen. Thus the slightly decreasing
trend for the set AMM0 becomes moderately increasing for the AMM1
one, and definitely increasing for the AMM2 case. For the $d$
flavor instead, the influence of the AMM is only light at $R=4$
and negligible at $R=7$. For most of the cases it does not have
the strength enough to change considerably the almost constant
behavior obtained for zero AMM.  The mass of the strange quark
does not show considerable variations, wherever the values of the
AMM. At medium densities $R=4$ it varies monotonously, while for
the higher density $R=7$ it exhibits fluctuations whose amplitude
increases with B,  but do not exceed $2$ MeV.

Finally, in Fig. 8 the magnetization as a function of the magnetic
intensity is shown for the fixed baryonic densities $R=4$ and
$R=7$. The typical oscillatory behavior is obtained, whose
amplitude as well as mean value increase with $B$.  It is
interesting to note that for the lower density the system  is
always paramagnetic, whereas for $n/n_0=7$ there is a change of
regime and for $B > 7.5 \times 10^{18}$ G becomes definitely
paramagnetic.

\section{Summary and Conclusions}

In this work a procedure to remove divergences in the $1/N_c$
approach to the SU(3) NJL model under a uniform magnetic field
$\bm{B}$ has been proposed. The calculations have been made by
using a covariant propagator for the quarks with constituent mass,
which takes account of the full effect of the magnetic field as
well as the effect of the anomalous magnetic moment. There are
divergencies which depend on the magnetic intensity. Since the
interaction used is an effective model of the strong interaction,
a full renormalization is meaningless. Therefore the divergent
terms are not ascribed to the renormalization of the external
magnetic field since, within the model used, it is not a dynamical
variable. In this work a systematic procedure to deal with such
kind of divergencies is proposed, instead. To obtain physically
meaningful results from the divergent contributions an analytical
regularization has been proposed which recovers the standard three
momentum cutoff scheme at $B=0$ and arbitrary baryonic density.
For this purpose a subtraction of fourth order in the vertices
$q_f B$ and $\kappa_f B$ is performed in the grand potential.
Since the regularization point is chosen at $B=0$ and fixed
baryonic density, the regularized quantities depend on the quark
masses $\tilde{M}_q$ evaluated in such conditions. The present
approach complements previous work, as for instance \cite{EBERT},
since it includes $\bm{B}$ dependent terms not considered before
as well as the additional coupling of the AMM of the quarks. The
regularization scale parameter, typical of the dimensional
regularization, has been chosen so as to maximize the agreement
with previous studies.

The regularized model has been used to study quark matter in
equilibrium against weak decay and electrically neutral, as can be
found in the composition of magnetars. A range of magnetic
intensities $10^{15} \text{G} \leq B \leq 10^{19} \text{G}$ and
baryonic densities $n \leq 1\, \text{fm}^{-3}$ have been analyzed,
which are adequate to describe such situation. A model for the
magnetic intensity in the interior of a magnetar \cite{PAL} has
been considered to test the results at finite density. For this
model the intensity $B$ is parameterized in terms of the baryonic
density and ranges between $10^{15} \text{G} \leq B \leq 5 \times
10^{18} \text{G}$.

The results at zero baryonic density have been compared with those
obtained with the commonly used prescription of \cite{EBERT}. In
general terms a qualitative agreement is obtained for low
intensities, but discrepancies become significative for strong
magnetic fields $B > 5 \times 10^{18} $ G. Hence one can conclude
that the study of magnetars will probably not evidence completely
these differences as in physical situations where the magnetic
field manifests with stronger intensity.

A contrast of the results with or without AMM shows that the
constituent mass of the $u$ flavor is the more sensitive quantity
to these effects, particularly in the medium to high density
regime.  The magnetization, instead, does not show clear evidence
of the influence of the AMM.

\begin{acknowledgements} This work has been partially supported by
a grant from the Consejo Nacional de Investigaciones Cientificas y
Tecnicas,  Argentina. \end{acknowledgements}

\appendix
\section{Regularization of the vacuum contribution to the thermodynamical potential}

In this section the Dirac sea contribution to Eq. (\ref{Kinetic})
is regularized, i.e. the contibution coming from the first term of
Eq. (\ref{DecomP}).

In momentum coordinates Eq. (\ref{Kinetic}) can be rewritten as
\begin{eqnarray}
\langle\bar{\psi}_f\,i
\gamma^0\partial_0\,\psi_f\rangle=-i\;\lim_{\epsilon\rightarrow
0^+}\,\int\, \frac{d^4p}{(2\pi)^4} e^{-i \epsilon p_0} p_0
\text{Tr}\left\{ \gamma^0 \,G_f(p)\right\} \label{EDenP}
\end{eqnarray}
Keeping only those terms corresponding to the Dirac sea it reduces
to
\begin{eqnarray}
-\frac{2\, i}{(2\pi)^4}{\sum_{n,s}}^\prime\frac{(-1)^n}{\Delta_n}
\int d^4p\,\frac{p_0^2\; e^{-p_\bot^2/\beta_f}}{u_p^2-(\Delta_n-s
K_f)^2+i \varepsilon} \left[\left(\Delta_n+s\,
M_f\right)L_n-\left(\Delta_n-s\, M_f\right)L_{n-1}\right]
\nonumber
\end{eqnarray}
where the argument of the Laguerre functions $L_k$ is
$2\,p_\bot^2/\beta_f$ and the primed sum has the same meaning as
in the main text.\\
As usual in analytic regularization, an undetermined scale factor
$\nu$ can be introduced \cite{RYDER}. After a Wick rotation in the
$p_0p_z$ space, the denominator in the previous equation can be
rewritten in exponential form by the well known procedure of
introducing a new integration on the variable $\tau$ which is well
defined in the Euclidean space
\begin{eqnarray}
\frac{2\nu^{-1}}{(2\pi)^4}
{\sum_{n,s}}^\prime\frac{(-1)^n}{\Delta_n} \int d^2p_\bot\,
d^2p_E\, p_4^2\, e^{-p_\bot^2/\beta_f} \left[\left(\Delta_n+s\,
M_f\right)L_n-\left(\Delta_n-s\, M_f\right)L_{n-1}\right]
\int_0^\infty d\tau \,e^{-\tau \left[p_E^2+(\Delta_n-s
K_f)^2\right]/\nu}\nonumber
\end{eqnarray}
where $d^2p_E=dp_4\,dp_z=d\theta dp_E p_E$ and $p_4=p_E \cos
\theta$. Changing the order, the integration over $d^2p_E$ can be
performed firstly obtaining $\pi \nu^2/2 \tau^2$. As the next step
one can integrate over $d^2p_\bot$ using polar coordinates and
with the help of formulae (7.414 6) of \cite{G&R}. Thus the
following expression is obtained
\begin{eqnarray}
\frac{\nu \beta_f}{8\pi^2}\int_0^\infty
\frac{d\tau}{\tau^2}\left[e^{-\tau(M_f- K_f)^2/\nu}+\sum_{n=1,s}
e^{-\tau (\Delta_n-s K_f)^2/\nu}\right]&=&\frac{\nu
\beta_f}{8\pi^2}\int_0^\infty
\frac{d\tau}{\tau^2}\left[\sum_{n=0,s} e^{-\tau (\Delta_n-s
K_f)^2/\nu}-e^{-\tau(M_f+ K_f)^2/\nu}\right]\nonumber\\
&=&\lim_{\epsilon \rightarrow 0} \frac{\nu
\beta_f}{8\pi^2}\int_0^\infty d\tau\, \tau^{\epsilon-2}
\left[\sum_{n,s} e^{-\tau (\Delta_n-s
K_f)^2/\nu}-e^{-\tau(m+K_f)^2/\nu}\right] \nonumber
\end{eqnarray}
In the last line a vanishing parameter $\epsilon$ has been
introduced in order to isolate the pole at $\tau=0$. By making a
trivial change of integration variable, but different for each
term between suare brackets one arrives to
\begin{eqnarray}
\frac{\nu \beta_f}{8\pi^2}\lim_{\epsilon\rightarrow
0}\left\{\sum_{s,n=0}\left[\frac{(\Delta_n-s
K_f)^2}{\nu}\right]^{1-\epsilon}-\left[\frac{(m+K_f)^2}{\nu}\right]^{1-\epsilon}\right\}\int_0^\infty
dt\, t^{\epsilon-2}\,e^{-t} \label{ToExpand}
\end{eqnarray}
The integral can be identified as $\Gamma(\epsilon-1)$. To put the
double summation in a simpler form, and bearing in mind that for
$\epsilon=0$ it reduces to
\begin{eqnarray}
\sum_{s,n=0}\frac{(\Delta_n-s
K_f)^2}{\nu}=\frac{1}{\nu}\sum_{s,n=0}\left(\Delta_n^2+
K_f^2\right)=\frac{2
\beta_f}{\nu}\sum_{s,n=0}\left(n+\frac{M_f^2+K_f^2}{2
\beta_f}\right)=\sum_s
\frac{2\beta_f}{\nu}\;\zeta\left(-1,\omega_f\right) \nonumber
\end{eqnarray}
where the series expansion for the zeta function was used, see for
instance Sec. 9.52 of Ref. \cite{G&R}, and
$\omega_f=(M_f^2+K_f^2)/(2 \beta_f)$. Thus, the following
approximation is proposed
\begin{eqnarray}
\sum_{s,n=0}\left[\frac{(\Delta_n-s
K_f)^2}{\nu}\right]^{1-\epsilon} \approx \sum_s\;\left(\frac{2
\beta_f}{\nu}\right)^{1-\epsilon}
\zeta\left(\epsilon-1,\omega_f\right) \label{Ansatze}
\end{eqnarray}
By insering Eq. (\ref{Ansatze}) into Eq. (\ref{ToExpand}) and
making a Laurent expansion around $\epsilon=0$ of the resulting
expression one obtains
\begin{eqnarray}
-\frac{1}{8\pi^2}\lim_{\epsilon \rightarrow
0}\Bigg\{\left(\frac{1}{\epsilon}+1-\gamma\right)\left[4\beta_f^2\zeta\left(-1,\omega_f\right)-\beta_f\left(M_f+
K_f\right)^2\right]+4 \beta_f^2 \frac{\partial}{\partial
z}\zeta\left(z=-1,\omega_f\right)\nonumber \\
 -4\beta_f^2 \zeta\left(-1,\omega_f\right)\,\ln\left(\frac{2\beta_f}{\nu}\right)+
\beta_f\left(M_f+K_f\right)^2
 \ln\left(\frac{\left(M_f+K_f\right)^2}{\nu}\right)+{\cal O}(\epsilon)\Bigg\}
 \label{SinglePole}
 \end{eqnarray}
Here a simple pole is evident, whose residue is  a polynomial of
fourth order in $B$. As $B \rightarrow 0$ this expression goes as
\begin{eqnarray}
\frac{m^4}{8\pi^2}\left(\frac{1}{\epsilon}+\frac{3}{2}-\gamma\right)
\label{EB0Lim}
\end{eqnarray}
that is, the typical behavior for a  Dirac particle is obtained.
The last divergence is usually tackled within this model by the
introduction of a constant 3-momentum cutoff $\Lambda$. In this
way the following finite contribution is assigned to it
\begin{eqnarray}
\langle\bar{\psi}_f\,i
\gamma^0\partial_0\,\psi_f\rangle_{NJL}=\frac{N_c}{8
\pi^2}\left[M_f^4
\ln\left(\frac{\Lambda+E_\Lambda}{M_f}\right)-\Lambda E_\Lambda
\left(\Lambda^2+E_\Lambda^2\right)\right], \label{Cutoff}
\end{eqnarray}
where $E_\Lambda=\sqrt{\Lambda^2+M_f^2}$.\\
In the following we apply to Eq. (\ref{SinglePole}) a procedure
that gets rid of the pole term and simultaneously ensures the
convergence to Eq. (\ref{Cutoff}) as $B\rightarrow 0$.

In Eq. (\ref{SinglePole}) it can be observed that the magnetic
dependence of the residue reduces to order two for $\kappa_f=0$,
hence the AMM is the cause of new divergencies depending on $B$.\\
It must be pointed out that this residue satisfies
\begin{eqnarray}F\left(M,\beta_f,K\right)&=&4\beta_f^2\zeta\left(-1,\frac{\lambda}{2
\beta_f}\right)-\beta_f\left(M+ K\right)^2 \nonumber \\
&=&\left(\sum_{k=0}^2 \sum_{j=0}^k \frac{\beta^{k-j}}{(k-j)!}
\frac{K^j}{j!} \frac{\partial^{k-j}}{\partial
\beta^{k-j}}\frac{\partial^j}{\partial
 K^j}+ \frac{K^4}{4!} \frac{\partial^4}{\partial
 K^4}\right) F(M,0,0)
\end{eqnarray}
Based on this feature the following  subtraction procedure is
proposed
\begin{eqnarray}
 \langle\bar{\psi}_f\,i
\gamma^0\partial_0\,\psi_f\rangle^{\text{D}}=\langle\bar{\psi}_f\,i
\gamma^0\partial_0\,\psi_f\rangle_{NJL}+\left[ 1-\sum_{k=0}^2
\sum_{j=0}^k \frac{\beta_f^{k-j}}{(k-j)!} \frac{K_f^j}{j!}
\left(\frac{\partial^{k-j}}{\partial
\beta_f^{k-j}}\frac{\partial^j}{\partial
 K_f^j}\right)_0- \frac{K^4}{4!} \left(\frac{\partial^4}{\partial
 K_f^4}\right)_0\right] \langle\bar{\psi}_f\,i
\gamma^0\partial_0\,\psi_f\rangle  \label{ERegProc}
\end{eqnarray}
Where the subindex $0$ means that the derivatives must be
evaluated at $\beta_f=0, K_f=0$. In this way the divergent term
cancels out and a finite contribution is generated as $\epsilon
\rightarrow 0$. The last one depends on the effective quark masses
$\tilde{M}$ evaluated at the regularization point. In the present
investigation this point is taken at a fixed baryonic number and
zero magnetic intensity. This notation  is used to  stress that
$\tilde{M}$ does not participate of the selfconsistent approach
defined by Eq.(\ref{EffMass}).
  For the properties of the derivatives of the
Hurwitz zeta function, see
for instance \cite{ELIZALDE}.\\
Thus, the final expression is
\begin{eqnarray}
\langle\bar{\psi}_f\,i
\gamma^0\partial_0\,\psi_f\rangle^{\text{D}}&=&\langle\bar{\psi}_f\,i
\gamma^0\partial_0\,\psi_f\rangle_{NJL}-\frac{N_c}{8\pi^2}\Bigg\{4
\beta_f^2 \frac{\partial}{\partial
z}\zeta\left(z=-1,\omega_f\right)
 +4\beta_f^2 \zeta\left(-1,\omega_f
 \right)\,\ln\left(\frac{\nu}{2\beta_f}\right)- 4\beta_f^2
 \zeta\left(-1,\tilde{\omega}_f\right)\ln\left(\frac{\nu}{\tilde{M}_f^2}\right)\nonumber \\
&&+ \beta_f\left(M_f+K_f\right)^2
 \ln\left(\frac{\left(M_f+K_f\right)^2}{\nu}\right)+\frac{\tilde{M}_f^4}{4}-
 \frac{K_f^4}{2}-\frac{\beta_f^2}{3}-2\beta_f K_f+\beta_f \left(\tilde{M}_f+K_f\right)^2 \ln\left(\frac{\nu}{\tilde{M}_f^2}\right)
 \Bigg\}
\end{eqnarray}
where $\tilde{\omega}_f$ is obtained from
$\omega_f$ by replacing $M_f$ by $\tilde{M}_f$\\
As explained in the main text, if $\nu=M_f^2$ is adopted the last
equation becomes
\begin{eqnarray}\langle\bar{\psi}_f\,i
\gamma^0\partial_0\,\psi_f\rangle^{\text{D}}&=&\langle\bar{\psi}_f\,i
\gamma^0\partial_0\,\psi_f\rangle_{NJL}-\frac{N_c}{8\pi^2}\Bigg\{4
\beta_f^2 \frac{\partial}{\partial
z}\zeta\left(z=-1,\omega_f\right)
 +4\beta_f^2 \zeta\left(-1,\omega_f
 \right)\,\ln\left(\frac{M_f^2}{2\beta_f}\right)- 8 \beta_f^2
 \zeta\left(-1,\tilde{\omega}_f\right)\ln\left(\frac{M_f}{\tilde{M}_f}\right)\nonumber \\
&&+2 \beta_f\left(M_f+K_f\right)^2
 \ln\left(1+\frac{K_f}{M_f}\right)+\frac{\tilde{M}_f^4}{4}-
 \frac{K_f^4}{2}-\frac{\beta_f^2}{3}-2\beta_f K_f+2 \beta_f \left(\tilde{M}_f+K_f\right)^2 \ln\left(\frac{M_f}{\tilde{M}_f}\right)
 \Bigg\}\label{DiracEnergy}
\end{eqnarray}

Using Eqs. (\ref{GrandPotential}),(\ref{FermiEnergy}) and
(\ref{DiracEnergy}) one obtains the regularized thermodynamic
potential by taking
\[
\langle\bar{\psi}_f\,i
\gamma^0\partial_0\,\psi_f\rangle=\langle\bar{\psi}_f\,i
\gamma^0\partial_0\,\psi_f\rangle^{\text{D}}+\langle\bar{\psi}_f\,i
\gamma^0\partial_0\,\psi_f\rangle^{\text{F}}.
\]
Furthermore, the quark condensates are evaluated within the
linearized approach \cite{KLEVANSKY, KUNIHIRO, VOGL, REHBERG}
simply as given by Eq. (\ref{Cond})
\[
\langle\bar{\psi}_f\psi_f\rangle=
\langle\bar{\psi}_f\psi_f\rangle^{\text{F}}+\langle\bar{\psi}_f\psi_f\rangle^{\text{D}}\]
The last term is evaluated by following the same steps as
described above. Thus one obtains
\begin{eqnarray}\langle\bar{\psi}_f\psi_f\rangle=
-\frac{N_c}{4\pi^2}&&\Bigg\{\left(\frac{1}{\epsilon}+\gamma\right)\left(\beta_f
K_f-\frac{M_f^3}{2}\right) +2\beta_f M_f
\ln\left(\frac{\Gamma(\omega_f)}{{\sqrt2
\pi}}\right)-M_f\left(M_f^2-\beta_f\right) \ln\left(\frac{M_f^2}{2
\beta_f}\right)\nonumber\\
&& +2 \beta_f\left(M_f+K_f\right)
\ln\left(1+\frac{K_f}{M_f}\right) + O(\epsilon)\Bigg\}
\end{eqnarray}
To get rid of the divergent term, the following subtraction scheme
is performed before taking the limit $\epsilon \rightarrow 0$
\[
\langle\bar{\psi}_f\psi_f\rangle^{\text{D}}=\langle\bar{\psi}_f\psi_f\rangle_{NJL}
+\langle\bar{\psi}_f\,\psi_f\rangle-\left[ 1+\beta_f K_f
\frac{\partial^2}{\partial \beta_f \partial
 K_f}\right] \langle\bar{\psi}_f\,\psi_f\rangle_0 \]
 where
\[\langle\bar{\psi}_f\psi_f\rangle_{NJL}=\frac{N_c}{2
\pi^2}M_f\left[M_f^2
\ln\left(\frac{\Lambda+E_\Lambda}{M_f}\right)- \Lambda
E_\Lambda\right]
\]
 The following result is obtained finally
\begin{eqnarray}
\langle\bar{\psi}_f\psi_f\rangle^{\text{D}}=\langle\bar{\psi}_f\psi_f\rangle_{NJL}
-\frac{N_c}{4\pi^2}&&\Bigg\{ 2\beta_f M_f
\ln\left(\frac{\Gamma(\omega_f)}{{\sqrt2
\pi}}\right)-M_f\left(M_f^2-\beta_f\right) \ln\left(\frac{M_f^2}{2
\beta_f}\right)\nonumber
\\&&+2 \beta_f\left(M_f+K_f\right)
\ln\left(1+\frac{K_f}{M_f}\right)+\tilde{M}_f^3 \Bigg\}
\end{eqnarray}


\newpage
\begin{figure}
 \includegraphics[width=0.9\textwidth]{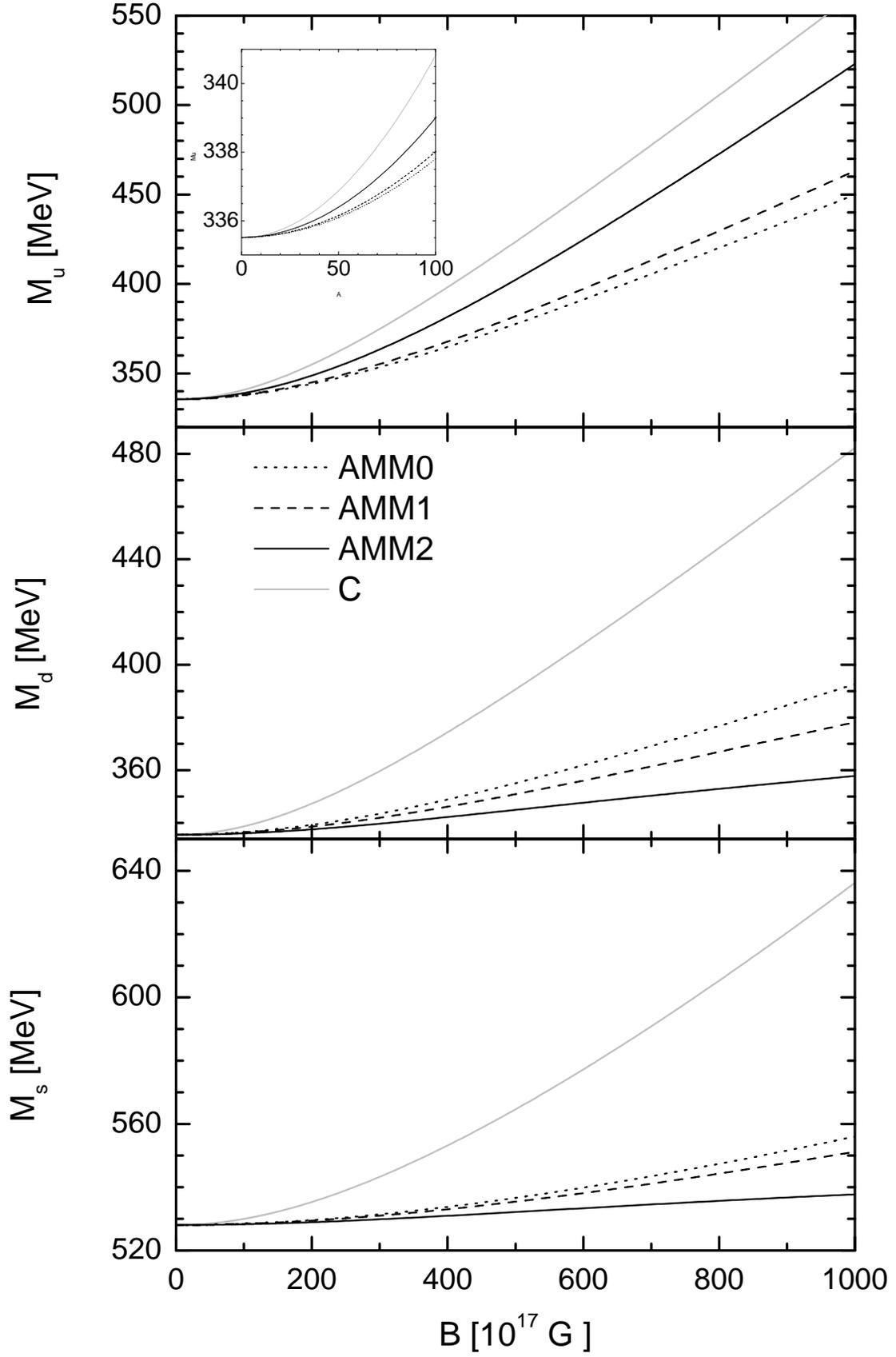}
 \caption{ The constituent quark mass as a function of the magnetic
 intensity at zero baryonic density. Two different regularization
 schemes, and results with or without AMM are compared as explained
 in the main text. In the case of the $u$ flavor an insertion shows
 details for the restricted range $B< 10^{19}$ G. }
 \end{figure}

\newpage
\begin{figure}
\includegraphics[width=0.9\textwidth]{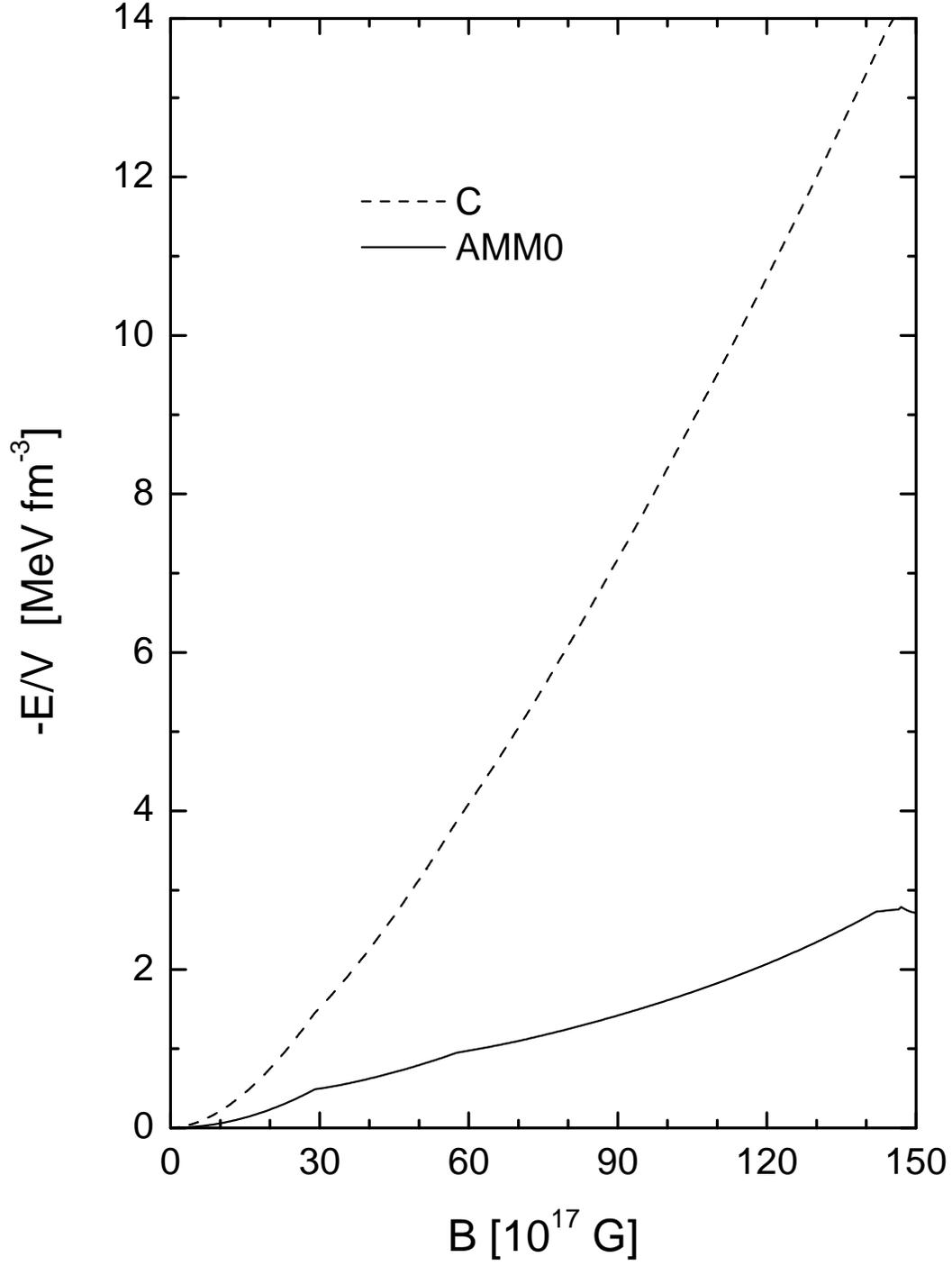}
\caption{ The energy density (with changed sign) at zero chemical
potential  as a function of the magnetic intensity for different
 regularization schemes. }
\end{figure}

 \newpage
 \begin{figure}
 \includegraphics[width=0.9\textwidth]{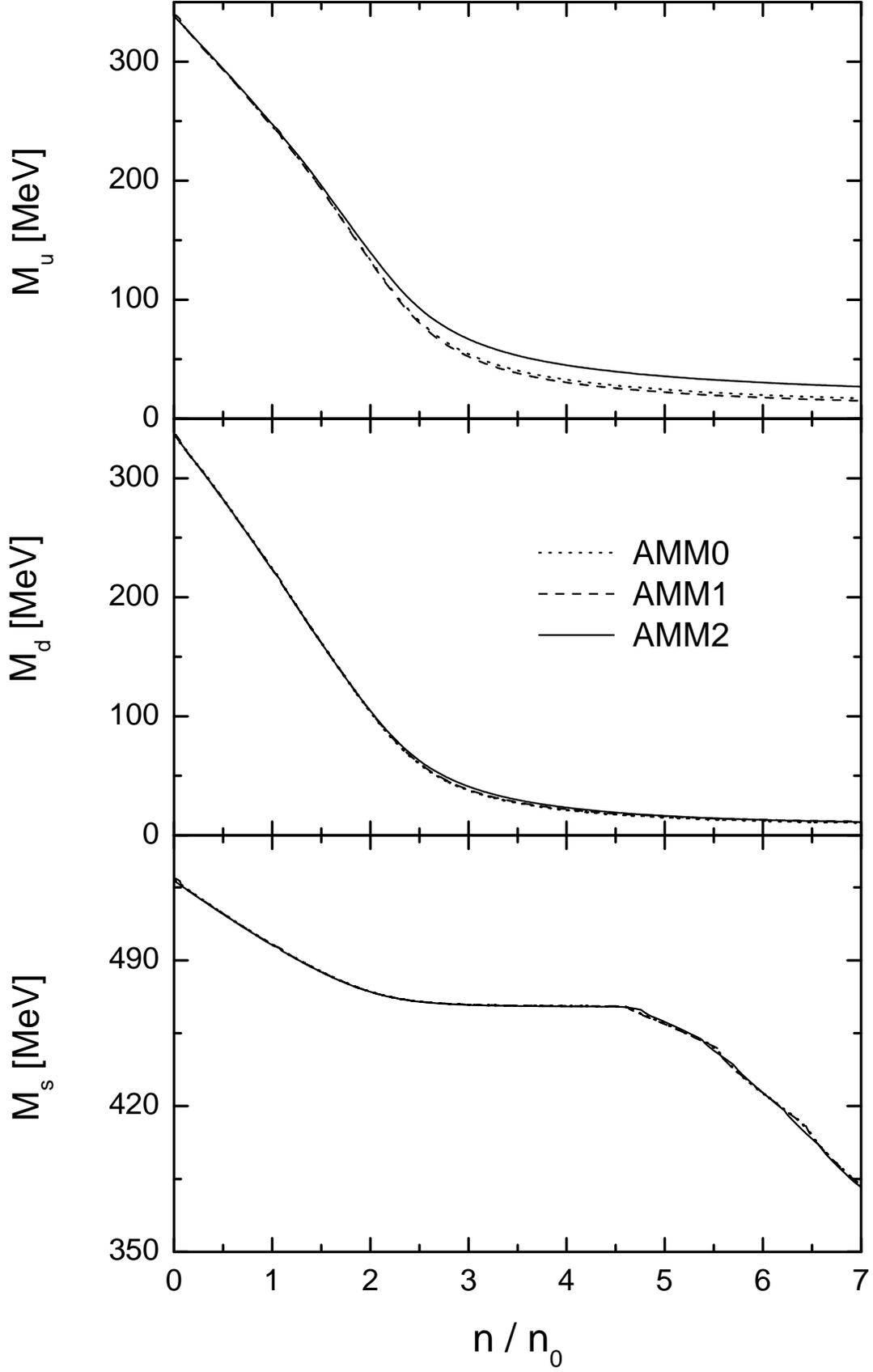}
 \caption{ The constituent quark mass as a function of the baryonic
 density for a fixed magnetic intensity $B= 10^{19}$ G. }
 \end{figure}

\newpage
\begin{figure}
\includegraphics[width=0.9\textwidth]{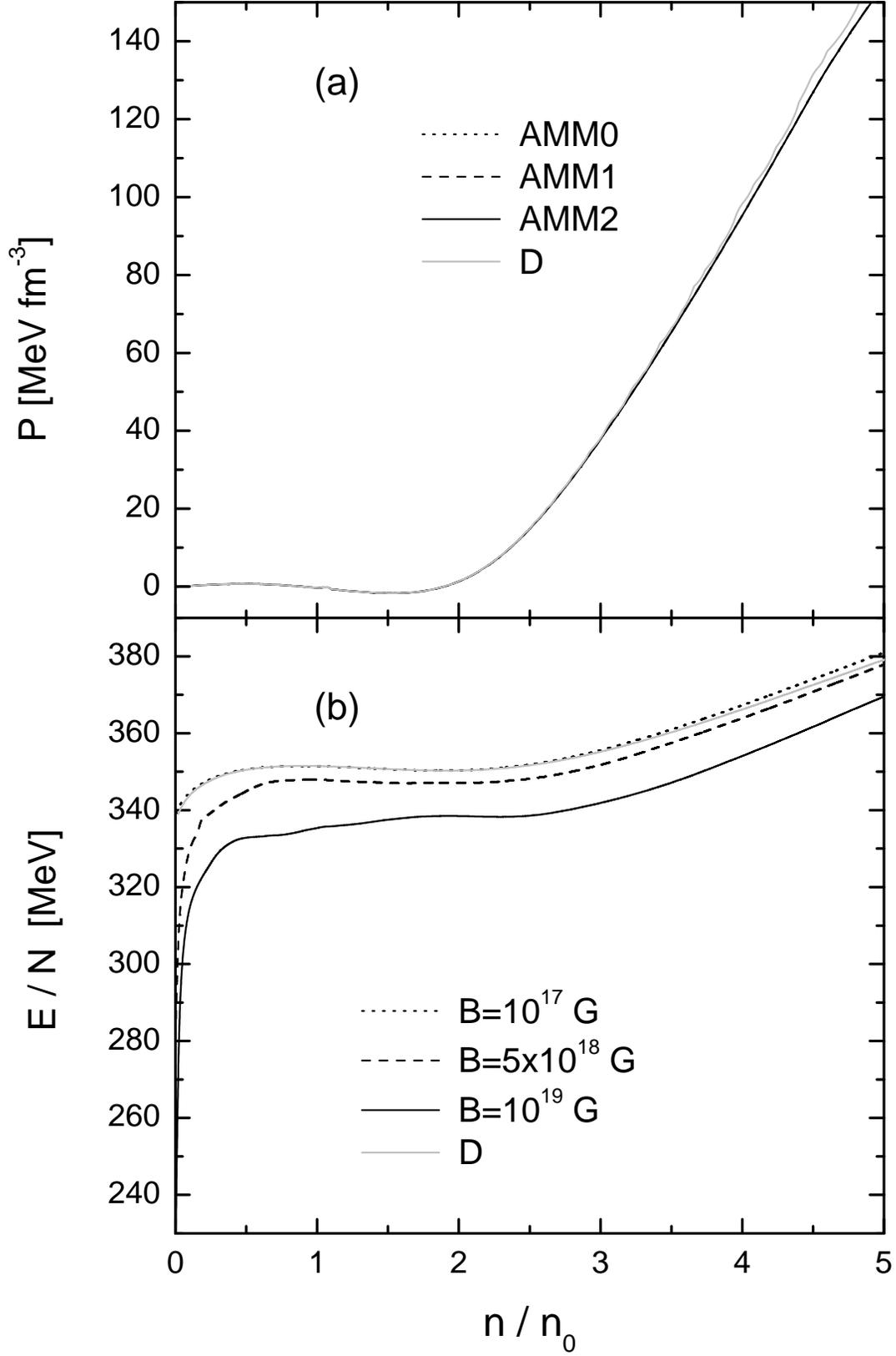}
\caption{ The thermodynamical pressure (a) and the energy per
particle (b)as functions of the baryonic density.  Results for a
model of the variation of the magnetic intensity in the interior
of a magnetar (case D) and  the results  at constant $B=10^{17}$ G
with the three set of AMM are included in the first case. In the
lower panel the case D is compared with results corresponding to
the set AMM1 and different magnetic intensities. }
\end{figure}

\newpage
\begin{figure}
\includegraphics[width=0.9\textwidth]{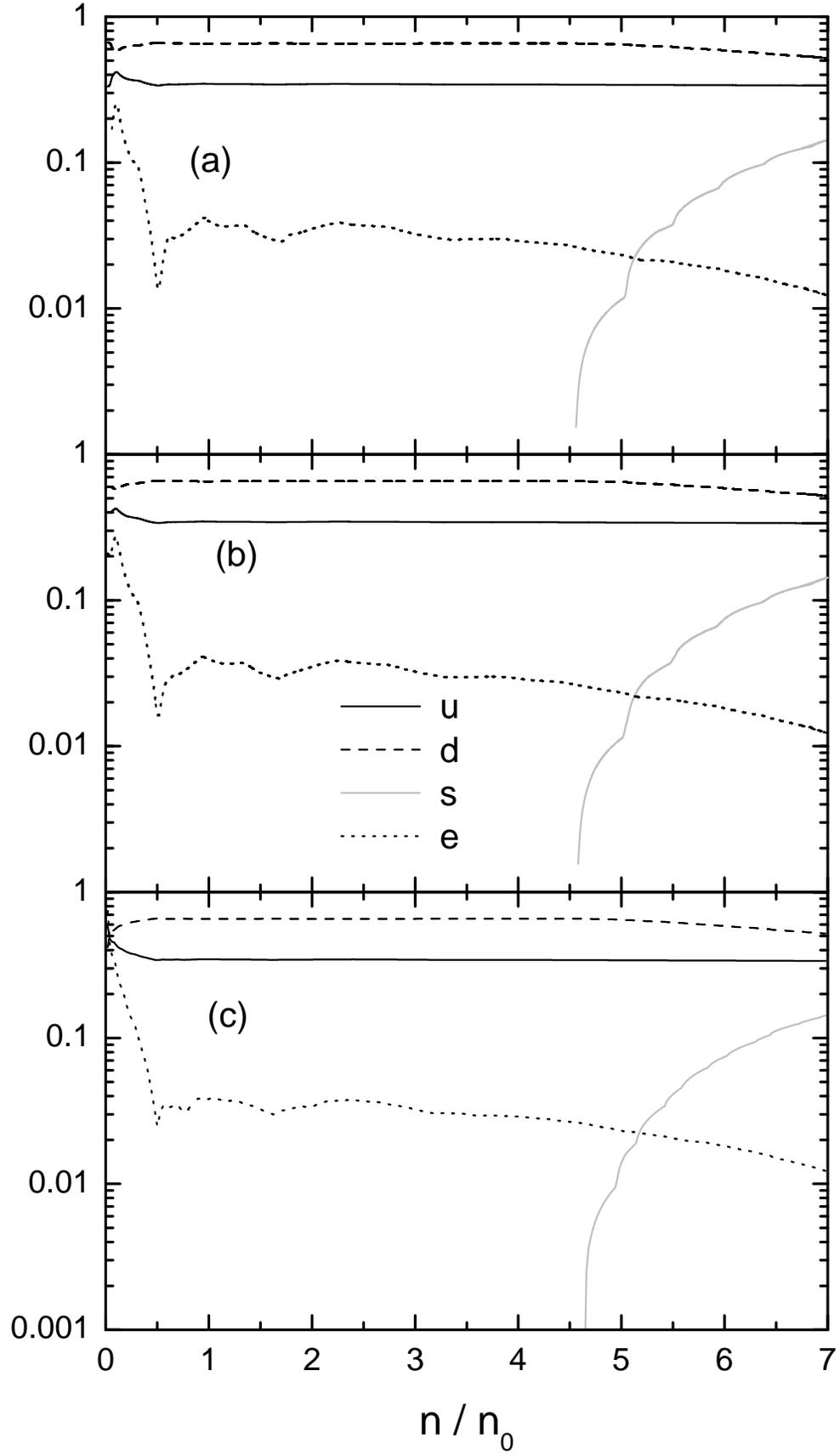}
\caption{ The relative number of particles as a function of the
baryonic density for a fixed magnetic intensity $B=5\times
10^{18}$ G. Results corresponding to the AMM0, AMM1 and AMM2
parametrizations are represented in the panel (a), (b) and (c)
respectively. }
\end{figure}
\newpage

 \newpage
 \begin{figure}
 \includegraphics[width=0.9\textwidth]{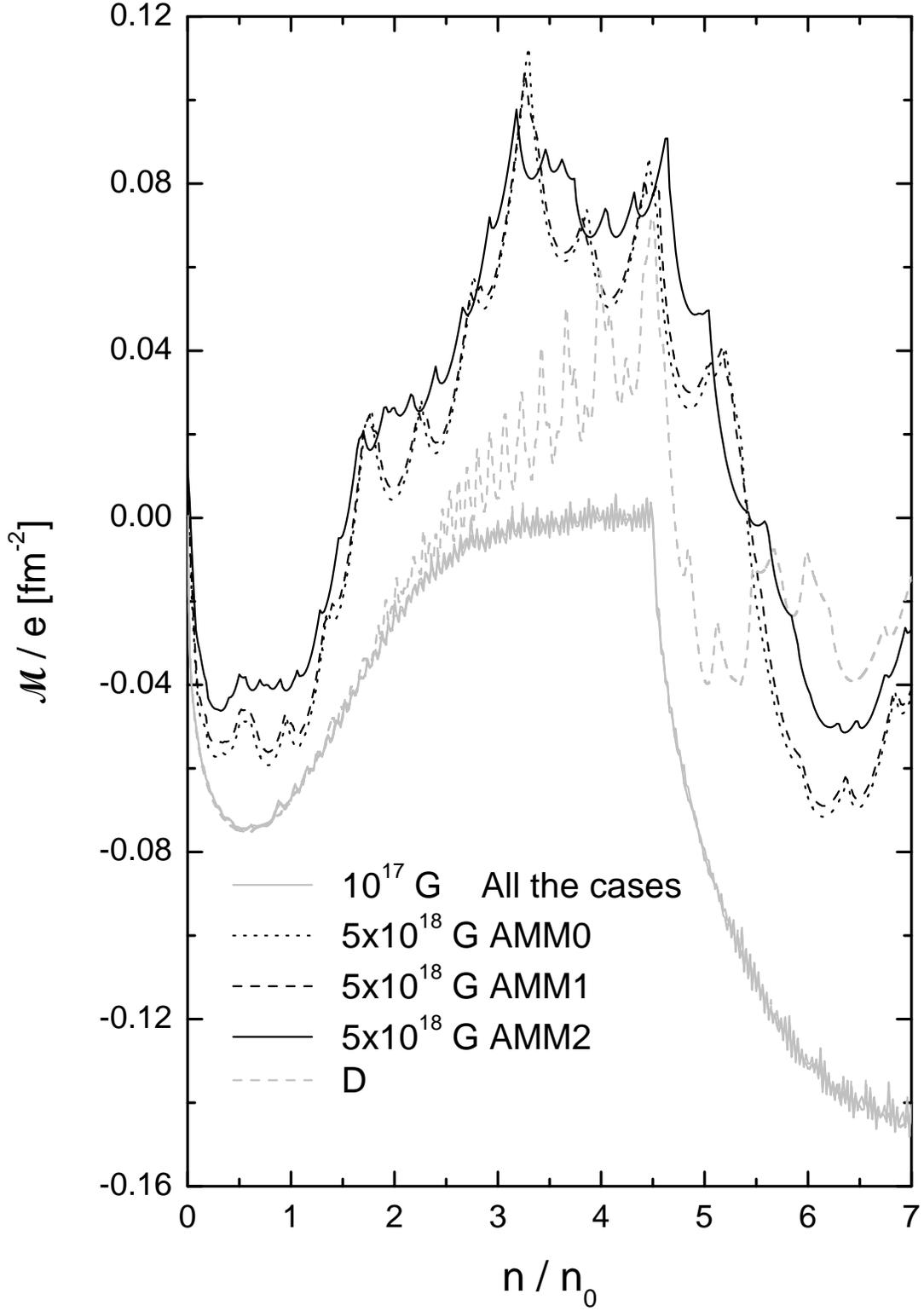}
 \caption{ The magnetization as a function of the baryonic density
 for fixed magnetic intensities $B=10^{17}$ and $B=5\times 10^{18}$
 G according to the convention shown. The results for variable
 intensity of Eq. (\ref{BModel}) are also shown (case D). }
 \end{figure}

 \newpage
 \begin{figure}
 \includegraphics[width=0.9\textwidth]{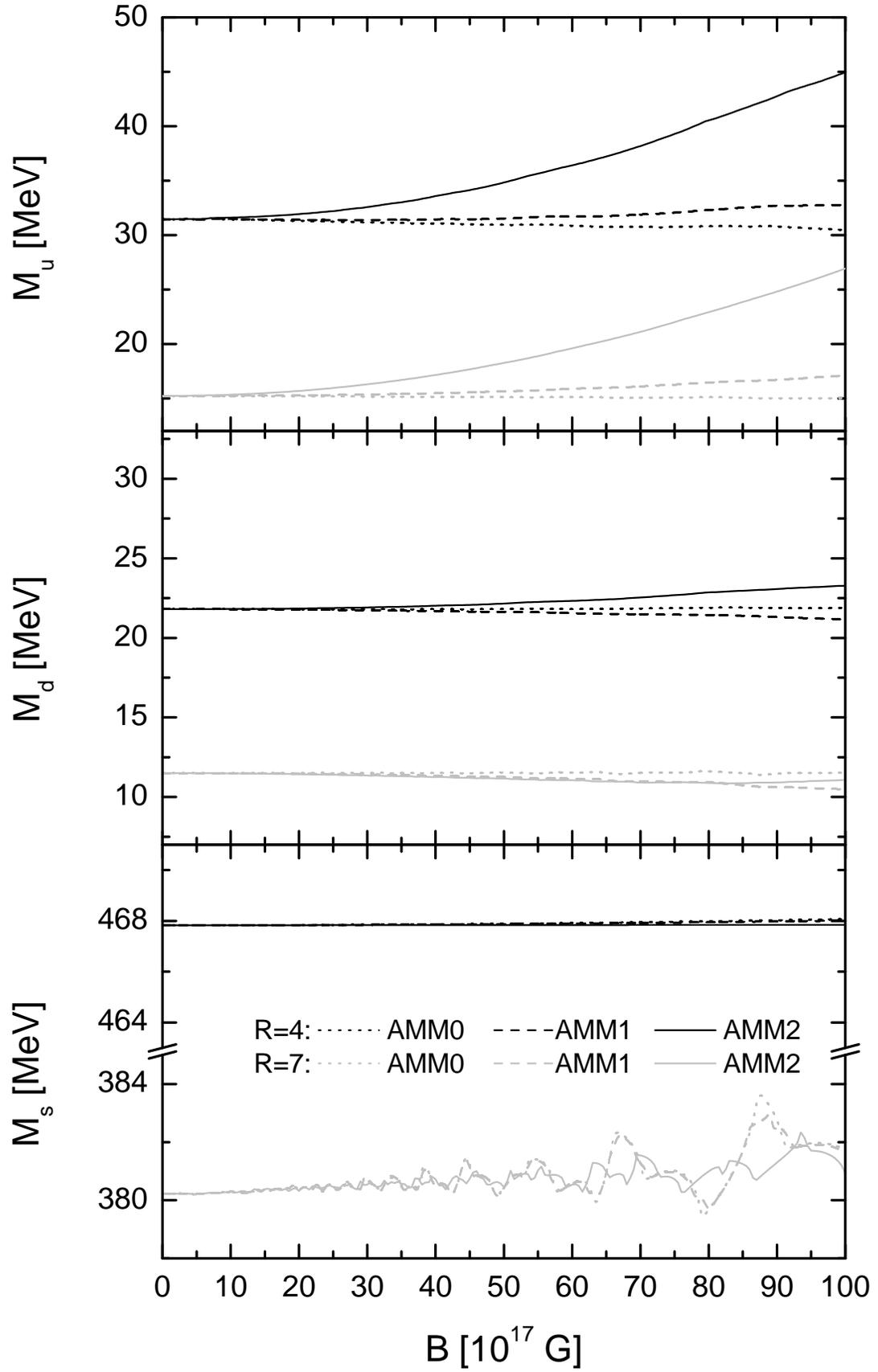}
 \caption{ The constituent quark mass as a function of the magnetic
 intensity for fixed baryonic densities $n/n_0=4$ and $7$. }
 \end{figure}

 \newpage
 \begin{figure}
 \includegraphics[width=0.9\textwidth]{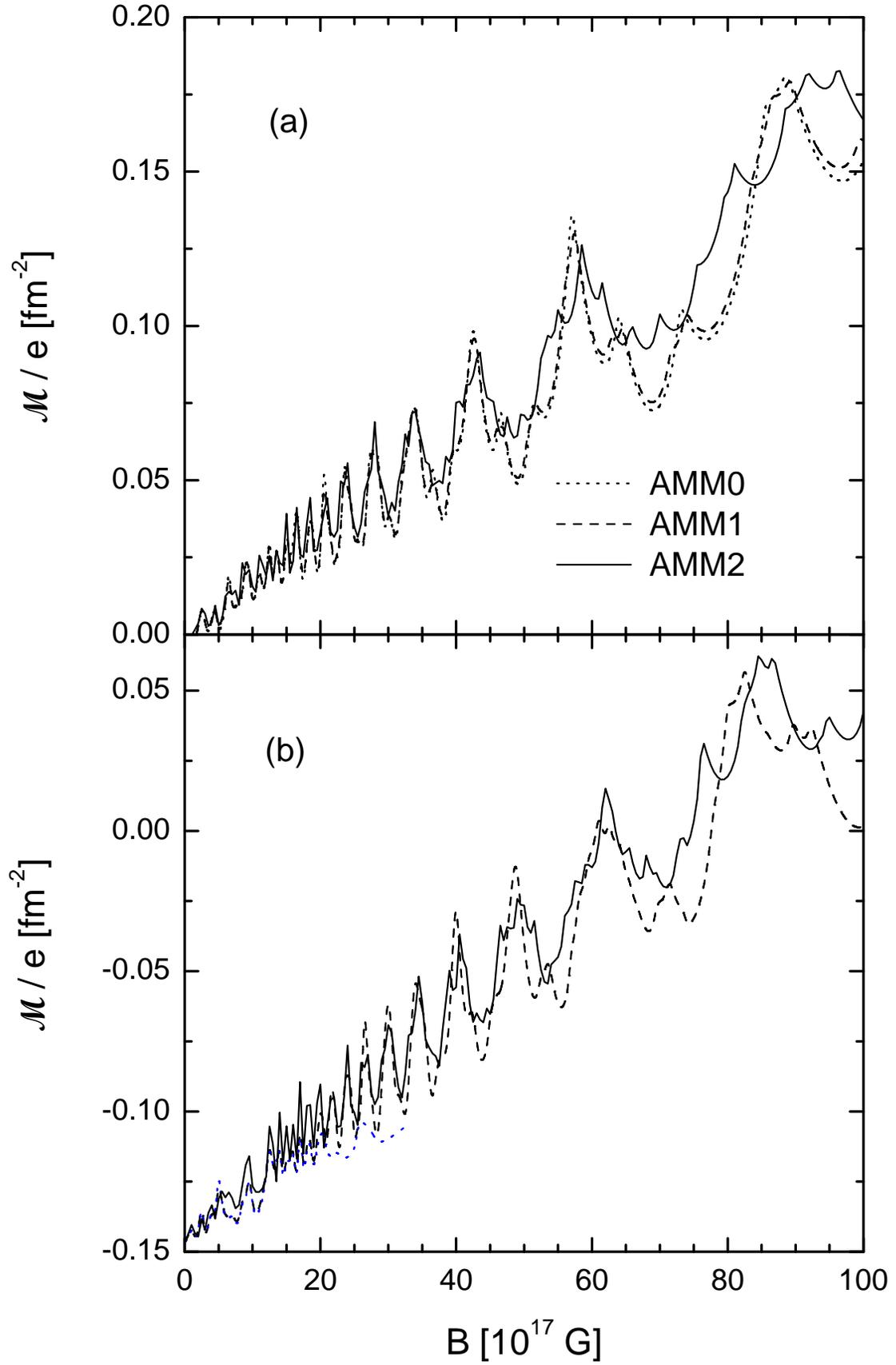}
 \caption{ The magnetization as a function of the magnetic
 intensity for fixed baryonic densities $n/n_0=4$ (a), and
 $n/n_0=7$ (b). }
 \end{figure}

\end{document}